# Enhancing Source Code Security with LLMs: Demystifying The Challenges and Generating Reliable Repairs

Nafis Tanveer Islam[*†], Joseph Khoury[‡], Andrew Seong[*†], Elias Bou-Hard[‡] and Peyman Najafirad[*†]

[*]Secure AI an Autonomy Laboratory
[†]University of Texas at San Antonio
[‡]Louisiana State University

*Abstract*—With the recent unprecedented advancements in Artificial Intelligence (AI) computing, progress in Large Language Models (LLMs) is accelerating rapidly, presenting challenges in establishing clear guidelines, particularly in the field of security. Numerous academic and industry efforts have focused on exploring Large Language Models (LLMs) to reliably identify, reason about, and address security vulnerabilities. However, current benchmarks indicate that state-of-the-art LLMs still face challenges in reasoning and require further research to overcome their limitations. That being said, we thoroughly identify and describe three main technical challenges in the security and software engineering literature that spans the entire LLM workflow, namely; *(i)* Data Collection and Labeling; *(ii)* System Design and Learning; and *(iii)* Performance Evaluation. Building upon these challenges, this paper introduces `SecRepair`, an instruction-based LLM system designed to reliably *identify*, *describe*, and automatically *repair* vulnerable source code. Our system is accompanied by a list of actionable guides on *(i)* Data Preparation and Augmentation Techniques; *(ii)* Selecting and Adapting state-of-the-art LLM Models; *(iii)* Evaluation Procedures. `SecRepair` uses a reinforcement learning-based fine-tuning with a semantic reward that caters to the functionality and security aspects of the generated code. Our empirical analysis shows that `SecRepair` achieves a *12%* improvement in security code repair compared to other LLMs when trained using reinforcement learning. Furthermore, we demonstrate the capabilities of `SecRepair` in generating reliable, functional, and compilable security code repairs against real-world test cases using automated evaluation metrics. Our data and source code are made publicly accessible here[1].

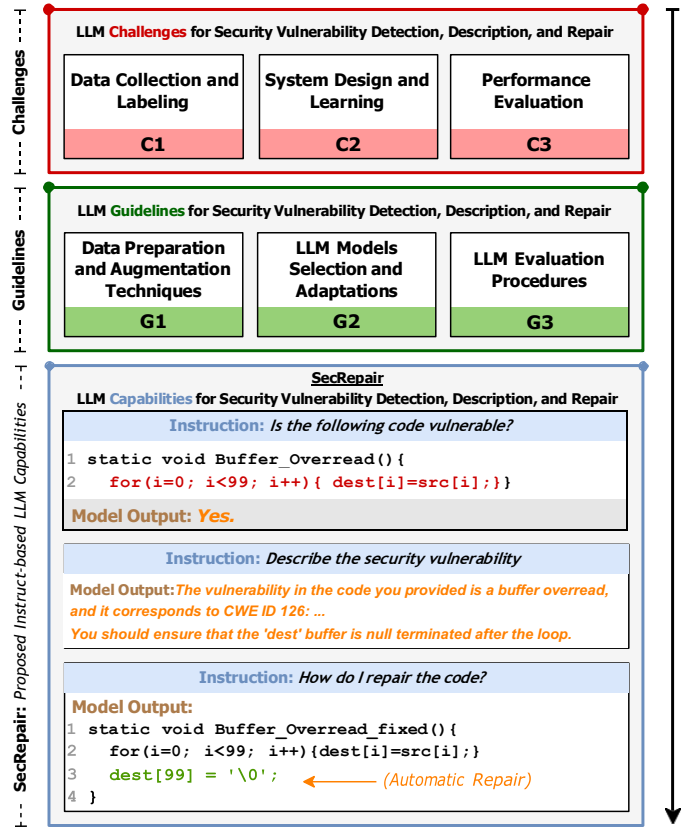

Fig. 1: **SecRepair Capabilities:** An overview of our thorough challenge identification and proposed actionable guidelines to build reliable instruct-based LLM capabilities for *detecting*, *describing*, and *repairing* code security vulnerabilities.

## I. INTRODUCTION

Recently, there has been a remarkable leap in Artificial Intelligence (AI) computing, propelled by industry giants such as NVIDIA, accompanied by notable progress in the research and development of Large Language Models (LLMs). These models are garnering attention across diverse domains, spearheading advancements in tasks such as generation (e.g., [1], [2], [3]), automation (e.g., [4], [5]), and commercial code auto-completion tools (e.g., `GitHub Copilot` [6], `TabNine` [7], `IntelliCode` [8], and `ChatGPT` [9]), to name a few.

In the intersection of software engineering and cybersecurity, LLM-based models are increasingly being utilized to identify vulnerabilities in critical and dependable software across high-stakes sectors, including vital government systems and healthcare infrastructures. This trend is underscored by initiatives such as the White House Executive Order on AI [10] and the ongoing 2024 DARPA AIxCC challenge [11]. Indeed, these systems necessitate unwavering vigilance and proactive measures to address potential code vulnerabilities and security risks [12], [13]. Even seemingly minor code vulnerabilities such as the Null Pointer Exception, commonly known as the *'Billion Dollar Mistake'* [14], can have far-reaching consequences by introducing security weaknesses and gaps. These vulnerabilities continue to be exploited by cybercriminals to propagate malware and engage in nefarious activities [15]. On a different note, with the extensive adoption of LLMs in software development, it has been shown that LLMs tend to generate *10%* more vulnerable code compared to a human developer [16]. This finding underscores the additional limitations and challenges we face in ensuring the development of secure code.

---

[1]https://anonymous.4open.science/r/Repair_Repo/

**Limitations of Existing Techniques and Critical Research Problems and Directions.** Initially, pioneering efforts including `Re-DeBug` [17], `VUDDY` [18], `MVP` [19], and `Movery` [20] have focused on identifying Vulnerable Code Clones (VCC) using parsing, fingerprinting, and signature-based techniques. However, these efforts did not explore the potential for describing and repairing security vulnerabilities in code, primarily due to the limited advancements in AI computing and LLM technologies at that time. Recent endeavors (e.g., [21], [22], [23]) have underscored the advantages of leveraging pre-trained LLM models for automated detection and repair of security vulnerabilities. Furthermore, other research initiatives (e.g., `VulRepair` [24] and `AIBUGHUNTER` [25]) have made notable progress in vulnerability detection by utilizing pre-trained LLM models. However, these initiatives still encounter technical limitations in terms of their data and model pipelines. As a result, their capacity to offer comprehensive vulnerability analysis and customized security code repairs for software developers is hindered. Given its nascent stages, the application of state-of-the-art LLM models is currently under scrutiny for their reliability in identifying and reasoning about security vulnerabilities [26]. Nevertheless, the significant impact of code vulnerabilities and their potential ramifications underscores the critical need to assist software developers in mitigating these risks. This involves leveraging advancements in AI computing and LLM models, which exhibit considerable potential but currently lack comprehensive guidelines and understanding.

Drawing upon the systematic analysis conducted by *Arp et al.* on common pitfalls in the machine learning workflow within computer security [27], as well as our extensive investigation of *19* recent top-tier security and software engineering papers, we have identified key challenges that arise when integrating and applying LLMs for the purpose of detecting and repairing security vulnerabilities in source code. These challenges encompass the entire LLM workflow, beginning with the data collection phase and extending through the subsequent phases, including model training, evaluation, and ultimately, deployment.

*Firstly*, defining an accurate threat model, along with incorporating relevant real-world Common Weakness Enumeration (CWEs), is crucial for establishing the objectives of the LLM pipeline development. Additionally, much of the recent research fails to clearly define how the developed LLM capabilities will be utilized, identify their intended audiences, or specify the types of attacks they aim to address.

*Secondly*, a prominent challenge in this domain is the absence or improper utilization of real-world vulnerability datasets, which serve as a fundamental pillar for the development of capable LLM models. Existing datasets in this field either rely on AI-generated data (e.g., [28]), have limitations in terms of instruction capabilities (e.g., [29]), or are language-dependent (e.g., `REEF` [30]).

*Thirdly*, another significant hurdle pertains to the training/learning and performance evaluation stages. As LLM models continue to proliferate, it becomes crucial to carefully consider model selection, training methodologies, and performance evaluation techniques. This ensures the accurate identification of vulnerabilities, precise generation of vulnerability descriptions, and reliable production of functional and flawless security repairs.

**Our Contribution.** Given the high stakes and the absence of clear guidelines, we aim to: *(i)* demystify the technical LLM workflow challenges encountered in the literature. We brake it down into three main categories; *data collection and labeling*; *system design and learning*; and *performance evaluation*. We provide a thorough discussion of each category and highlight specific security limitations associated with each challenge; *(ii)* propose `SecRepair`, an instruction-based LLM that utilizes reinforcement learning-based fine-tuning. It incorporates a semantic reward system to enable the *identification*, *description*, and *repair* of code security vulnerabilities. We accompanied `SecRepair` with a set of guidelines that serve as actionable recommendations to enhance the security posture of source code with LLMs; *(iii)* evaluate `SecRepair` with real-world datasets. Figure 1 provides a holistic view of the challenges, guidelines, and highlights on the instruction based capabilities of `SecRepair`.

In summary, the contributions of this paper are as follows:

- **Threat Model Devising.** We devise a threat model that precisely outlines the goals of LLMs in enhancing source code security. Particularly, we focus on real-world vulnerability outbreaks, which offer valuable insights into the usage, target audience, and types of attacks that need to be addressed (check §II).

- **Challenge Identification.** By following a systematic analysis proposed in [27] and thoroughly reviewing *19* top-tier security and software engineering papers, we identify key challenges and limitations in the literature. These challenges span over all stages of the LLM workflow and have serious security implications, impacting the reliability of LLMs in reasoning about security vulnerabilities (check §III).

- **Guideline Formulation.** We formulate an exhaustive list of actionable guidelines to correctly employ and augment existing vulnerability datasets and carefully select and adapt state-of-the-art LLMs to enhance source code security (check §IV).

- **SecRepair.** Based on our established guidelines, we introduce `SecRepair` an LLM for the purpose of identifying, describing, and providing security repairs. Our results demonstrate that `SecRepair` achieves a significant improvement of *12%* in Cosine similarity and CodeBERTScore values for vulnerability repair when using reinforcement learning compared to supervised fine-tuning.

The remaining sections of the paper are organized as follows. In §II, we devise our threat model. In §III, we identify and thoroughly discuss the literature challenges. In §IV we present `SecRepair` along with a comprehensive list of actionable guidelines. Following this, in §V, we outline the experimentation conducted and showcase the capabilities of `SecRepair`. In §VI, we present an ablation study. In §VII we discuss the related work. Finally, our work is concluded in §VIII.

## II. THREAT MODEL

Recently, a security vulnerability known as `CVE-2024-3094` *(i.e., CVSS score = 10.0 - critical)* was discovered in the open-source library XZ Utils [31]. This vulnerability, through a backdoor, enables remote code execution (RCE), which resulted from a rogue maintainer introducing malicious code disguised as test files into the library [31], [32], [33]. By going unnoticed, this vulnerability could have had serious consequences on a significant number of servers on the internet, potentially leading to unauthorized access, data breaches, and other malicious activities.

On another hand, the uncontrollable rise of neural language modeling and state-of-the-art Large Language Model (LLM)-based auto-completion tools (e.g., `GitHub Copilot` [6], `TabNine` [7], `IntelliCode` [8], and `ChatGPT` [9]) has introduced a new set of deficiencies, primarily stemming from generating weak programming implementations, further compounding the challenge of ensuring secure and reliable software [34], [35], [36], [37].

Accordingly, our threat model focuses on the code injection attack, which consists of injecting code that is then interpreted/executed by the application [38]. These attacks originates from *(i) a potential cyber threat actor, as seen in the case of the XZ vulnerability outbreak, and (ii) an inexperienced software developer who may inadvertently introduce vulnerabilities while using auto-completion tools or making mistakes.* Figure 2, step one illustrates the code injection stage, while step two portrays a malicious event that affects both the data and the system following the execution of the vulnerable software.

To this end, our work is centered around empowering software developers and security experts with reliable and trustworthy LLM capabilities to identify, describe, and automate security vulnerability repair. Figure 2, step three, illustrates the interaction between a software developer or a security expert and an instruction-based LLM model. In this interaction, the user provides a source code as input and specifies an instruction to either describe or repair a vulnerability (in this case a 'repair' instruction). A reliable and capable LLM model is expected to fulfill these instructions by providing an automated security code repair that is both functional and compilable in the event of vulnerability detection. Check Figure 1 for the full capability presentation of our proposed system, namely, `SecRepair`.

Our threat model primarily focuses on code-based vulnerabilities that occur at the function level of a program, namely, intra-procedural analysis. The vulnerability analysis examined in this paper covers various types of CWEs. Firstly, we address vulnerabilities that can lead to program crashes by inducing memory leaks or buffer overflows, including `CWE-787`: *Out-of-bounds Write*; `CWE-121`: *Stack-based Buffer Overflow*; and `CWE-119`: *Improper Restriction of Operations within the Bounds of a Memory Buffer*. Secondly, we focus on vulnerabilities that do not crash the program but instead enable threat actors to gain control of a system through code or malware injection, illegal memory access, and obtaining root access to a system. These include; `CWE-89`: *SQL Injection*, `CWE-94`: *Improper Control of Generation of Code*, and `CWE-20`: *Improper Input Validation*. Furthermore, our proposed system enhance developer knowledge by providing descriptions of vulnerabilities along with their corresponding CWE numbers.

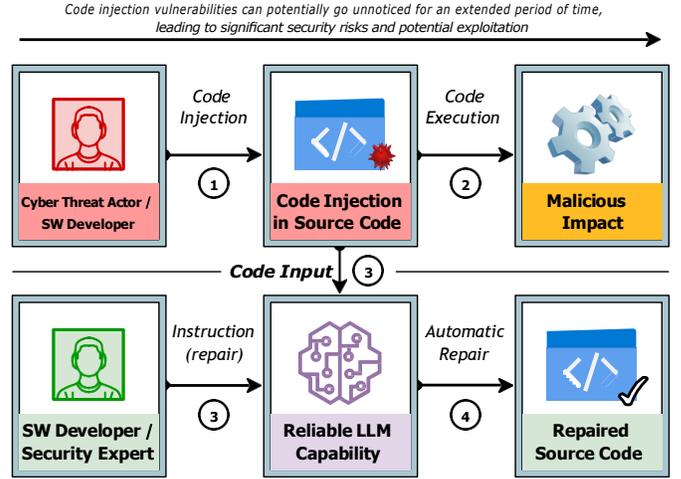

Fig. 2: **Threat Model:** A depiction of a *code injection* threat model involves a cyber threat actor leveraging one of the top CWEs to inject malicious code into an open-source software (OSS) project, with the goal of compromising its integrity and gaining unauthorized system access or control. *Our goal is to equip software developers and security experts with well-developed and reliable LLM capabilities for automated security code repair.*

## III. CHALLENGES IN SOURCE CODE SECURITY ANALYSIS USING LLM MODELS

Despite the considerable success and unprecedented capabilities demonstrated across diverse domains, the application of LLM models is still in its nascent stages. Further research and development are imperative to establish their reliability, especially in the context of source code vulnerability detection and repair. Notably, this aspect has been extensively studied and evaluated by *Ullah et al.* [26], which encompasses multiple vulnerability datasets, state-of-the-art LLM models, and real-world CWE test cases. In this section, we will discuss three primary technical challenges with their respective security implications associated with the research and development of LLM pipelines (i.e., datasets, models, and evaluations) for detecting and repairing security vulnerabilities in source code. Our identification of these challenges is based on the analysis of learning-based security systems presented in [27], and involves extracting deficiencies and weak points rooted in top-tier security and software engineering papers on the topic.

> **C1 - LLM Challenges in Data Collection and Labeling for Security Vulnerability Detection, Description, and Repair**
>
> *Security Implications: Inaccurate identification/description of vulnerabilities, lack of objective reasoning in code analysis, biased distinctions between vulnerable/non-vulnerable samples, lack of relevance to important and emerging top CWEs.*

> **C2 - LLM Challenges in System Design and Learning for Security Vulnerability Detection, Description, and Repair**
>
> *Security Implications: Non-functional nor compilable generated security repair, snooping on synthetic vulnerable data.*

*Challenge 1 (C1): Data Collection and Labeling*
**Description.** Technically speaking, the trainable weights of LLM models are shaped by the data and fitted to represent specific relationships and contextual understanding accurately. Thus, the data serves as the foundation for the model's reasoning and generation of content. For optimal performance in a learning system, including application-specific LLM models, the utmost importance lies in the elements of data source, quality, quantity, relevance, label, and distribution. In the realm of security, ensuring these data properties become even more complex due to the scarcity of data caused by privacy and security concerns, as well as the manual efforts required by experts for generation, collection, and labeling. As such, this will ultimately lead to data that suffers from *(i)* label inaccuracy; *(ii)* data leakage; and *(iii)* sampling bias. In studies such as [39] and [40], the determination of vulnerable and non-vulnerable code samples were not made by security experts. Instead, it relied on vulnerability keywords and regular expressions proposed by [41], which can not guarantee accurate labeling. Additionally, we observed data leakage in multiple vulnerability datasets (e.g., `MVD` [42], `Devign` [39] and `ReVeal` [40]) where the function name in the code samples leaked the CWE number, which reflects the label of the data instance. Other works, such as [40], suffer from a significant imbalance between vulnerable and non-vulnerable data points and real vs synthetically generated vulnerabilities [28].

**Security Implications.** Using LLM models trained on inaccurate data will undoubtedly result in inaccurately identifying vulnerabilities in input source code. In other words, encountering false positives and negatives when identifying vulnerabilities and classifying CWEs as demonstrated in [26]. Additionally, building upon data with leaked label will result into detrimental effect on the LLM model's ability to genuinely understand the code's objectives as demonstrated in [43]. Moreover, the challenge of imbalanced data can inherently lead to bias in determining whether a code sample is vulnerable or not. Not to mention the problem of synthetic vulnerable code, which imposes serious reasoning limitations on LLM models in practice when they are exposed to emerging top CWEs and real-world zero-day vulnerabilities that have never been seen before. To ensure a higher level of security in software systems, it is crucial to consider the security implications of all data-related challenges. This is particularly important when utilizing poorly designed and implemented LLM-based capabilities for security code identification and repair during the development of critical software systems.

*Challenge 2 (C2): System Design and Learning*
**Description.** Numerous LLM models are proliferating, each with its own unique design choices, technicalities, and intended usage. As such, selecting the correct LLM and ensuring a rigorous training process is paramount in building a reliable LLM for security vulnerability identification and repair. In the pursuit of enhancing source code security for software developers, it is vital to enable a chat-based assistant, namely, instruct-based LLM models that can reliably identify, describe, and autonomously repair vulnerabilities, as expounded upon in §II. Initially, pre-instruct LLMs, such as `T5` [44], `CodeT5` [45], `CodeT5+` [46], and `CodeRL` [47], solely relied on the input to generate responses, without the ability to respond to specific user instructions. In contrast, instruction-based generative models such as `GPT-4` [48], `Gemini` [49], and trainable models including `CodeLLaMa` [50], `CodeGen2` [51], `LLaMa 3` [52], and `Mistral` [53] are trained with instructions, making them more versatile for general-purpose use. However, the training methodologies employed by these models, such as supervised fine-tuning processes (e.g., `LLaMa` [52]) and supervised fine-tuning with reinforcement learning from human feedback (e.g., `InstructGPT` [54]), do not directly fit the down task of vulnerability detection, description, and repair. Without careful consideration, several recent works including `VulRepair` [24] and others (e.g., [55]) have predominately relied on supervised fine-tuning approaches that utilize cross-entropy loss as the sole criterion for training the model to generate the correct output. Moreover, the dearth of real-world vulnerabilities in comparison to synthetic vulnerability datasets as seen in [56], [26] and generated by SARD [57] and NIST [58] using various LLM models (e.g., GPT-4 [48], Falcon [59], among others) exacerbate the problem of data snooping manifested by training and testing data not available in practice.

**Security Implications.** The sole reliance on supervised fine-tuning in LLM models is fundamentally inadequate when dealing with programming languages that have specific syntax and semantics. This approach fails to address the potential generation of non-compilable code, functionally incorrect code, or code that is susceptible to security vulnerabilities as demonstrated in [60]. This focus on cross-entropy loss overlooks the necessity for syntactical, semantic, and functional accuracy, leading to unreliable and potentially dangerous LLM models in real-world applications. Furthermore, these models often lack robust mechanisms to ensure that the generated code adheres to best practices in security and functionality. The absence of comprehensive evaluation methods compounds this issue, as there is no rigorous assessment of the model's ability to generate secure, functional code. This glaring oversight highlights the urgent need for an integrated training process that seamlessly embeds syntactical, semantic, and functional information, ensuring that LLMs can produce high-quality, secure code repair suitable for real-world applications. Moreover, there is the issue of snooping on synthetic vulnerable data, which can lead to optimistic results. While synthetic vulnerabilities are generated based on known data distribution, they often fail in practice when confronted with previously unseen vulnerabilities.

*Challenge 3 (C3): Performance Evaluation*

> **C3 - LLM Challenges in Performance Evaluation for Security Vulnerability Detection, Description, and Repair**
>
> **Security Implications:** *Insecure and potentially non-functional generated security repairs due to the lack of robust performance metrics in the domain of LLMs for Security Vulnerability Detection, Description and Repair.*

**Description.** Measuring the performance of generative tasks, particularly when it comes to generating functional or security repairs for source code, presents significant challenges. Numerous efforts (e.g., [16], [23], [60], and others) have been undertaken to utilize state-of-the-art LLMs for tasks such as code security repair. However, the evaluation methods employed in these studies are still questionable and lack robustness. To illustrate, studies conducted by [16], [26], and [23] showcase the utilization of small sample sizes, typically comprising fewer than 500 instances, for the purpose of code security repair. This limited sample size has proven advantageous for facilitating manual evaluation. However, the case is different when dealing with large-scale repair tasks involving over 100,000 samples. As presented in [60], [55], [24], [21], [61], and [62], automatic evaluation becomes increasingly challenging. Two primary methods have been identified for evaluating such works. Firstly, when input test cases for a given function are available, work such as [61] utilizes the Pass@K metric proposed in [63]. Secondly, in the absence of test cases, studies such as [62], [24], and [55] resort to Exact Match, BLEU [64], and Rouge-L [65] scores for automated evaluation. Additionally, as part of the automated evaluation process, [26] employed Cosine Similarity and `GPT-4` as an LLM to evaluate generated outcomes.

**Security Implications.** Despite these efforts, these automated metrics are fundamentally inadequate. They offer a superficial measure of similarity but fail to guarantee the functionality or security correctness of the generated security code repair. Moreover, they do not ensure that the code is even compilable, a critical prerequisite for execution. This lack of reliable evaluation metrics leads to the risk of deploying non-functional or insecure code, undermining the entire purpose of generative models for code repair.

## IV. SECREPAIR: PROPOSED APPROACH VIA ACTIONABLE GUIDELINES

**Problem Formulation.** Each function is defined as $f$; if the function is vulnerable, it is defined as $f_{vul}$ and the corresponding ground truth repaired function as $f_{rep}$. The vulnerability description is denoted as $D$. The vulnerable function $f_{vul}$ is converted to tokens denoted as $T$. Our methodology utilizes a multitask approach encompassing vulnerability identification, repair, and description. Initially, it identifies vulnerable code $V$ as a binary classification task, followed by repairing the vulnerability $f_{rep}$ and, subsequently, generating a description $D$ of the vulnerability.

Let us consider the task of vulnerability identification and repair as our first step. For this set of tasks, A function $f$ could be vulnerable or non-vulnerable. It is used as the input to the model with an instruction to generate $V$ and $f_{rep}$ where $V$ is a binary variable indicating the existence of security vulnerability using the words "YES" and "NO" in the input function, and $f_{rep}$ is the repaired function. For the tasks of describing the vulnerability, we define $D$ as the vulnerability description. Following the instructions, our system generates $D$, where $D$ provides a detailed description of the identified vulnerability. When the task is to repair the vulnerability $f_{rep}$, the input is the vulnerable function $f_{vul}$ with instruction. Here, we denote $y$ as the generated by the model. The instruction changes based on whether we want to identify, describe, or repair the function. The capabilities in Figure 1 show the instruction, input, and output generated by the model.

In this study, we aim to address three research questions (RQs) regarding the effectiveness and capabilities of our proposed system, `SecRepair`.

**RQ1:** *Can we automatically identify code vulnerability of static source code?*

**RQ2:** *Can we comprehensively describe the code vulnerability of the vulnerable code*

**RQ3:** *Can we optimize using reinforcement learning and repair the security issue in a vulnerable code?*

By addressing these research questions, we aim to advance cybersecurity techniques by adhering to our proposed guidelines in the next Subsections. These guidelines ensure balanced and obfuscated data collection, aligning training data and metrics with real-world applications, and implementing rigorous evaluation to guarantee syntactic, functional, and security correctness. The guidelines are designed to mitigate vulnerabilities in code developed by human programmers.

**SecRepair.** In this section, we introduce `SecRepair`, a novel code repair system designed to address the critical challenges in static code vulnerability analysis. Our approach is underpinned by three fundamental Guidelines: G1:Data Preparation and Augmentation Techniques, G2: Selecting and Adapting Models, and G3: Evaluation Procedures. Figure 3 illustrates the architecture of our proposed work. The proposed architecture is divided into three stages: *(i)* Instruction Dataset Preparation, *(ii)* Code Vulnerability Identification and Description, and finally, *(iii)* Vulnerability Repair. In this section, we will show an in-depth implementation of the guidelines **G1**, **G2**, and **G3** to build our proposed system, namely, `SecRepair`.

*Guideline 1 (G1): Data Preparation and Augmentation Techniques*

> **G1 - LLM Guidelines in Data Preparation and Augmentation Techniques for Security Vulnerability Detection, Description, and Repair**

Our data collection strategy prioritizes balanced data from diverse sources to mitigate issues such as Data Snooping, Label Inaccuracy, and Sampling Bias following **G1**. Additionally, we ensure the obfuscation of data to prevent unintended data leakage. This rigorous approach to data collection lays a solid foundation for the subsequent training and evaluation stages, guaranteeing a more robust and unbiased dataset.

This study introduces `InstructVul`, the first instruction-based dataset designed for a vulnerability identification and repair system, including vulnerability description.

`InstructVul` consists of three components corresponding to three tasks: *(i)* Vulnerability Identification, *(ii)* Vulnerability Description, and *(iii)* Vulnerability Repair. The subsequent subsections formally outline the `InstructVul` dataset, followed by a detailed description of its creation process.

**Formal Definition.** Each data entity contains three components: instruction, context input, and output. The instruction is denoted by *I*, which is a statement stating one of the three tasks demonstrated in Figure 1. The second component context input $C_i$ is a supplement for each instruction, which is either the vulnerable and non-vulnerable function denoted as $f_{vul}$, $f_{rep}$. The third component is the output $y$, the expected LLM-generated response.

```
FUNCTION_NAME_STRIP_QUERY: Final[str] = '''
(function_definition
  declarator: (function_declarator
    declarator: (identifier) @func-def
  )
)
(call_expression
  function: (identifier) @call-expr
)
'''

parser = Parser()
language = Language(str(LANGUAGES_FILE), 'cpp')
parser.set_language(language)

def strip_func(func_def: str) -> str:
    func_names = set(n.text.decode() for n, t in
    language.query(STRIP_QUERY)
              .captures(parser.parse(func_def
    .encode()).root_node))
    stripped_mappings = {n: f'func_{uuid.uuid4().hex
    [:16]}'
                for n in func_names}
    stripped_def = func_def
    for orig_name, strip_name in stripped_mappings.
    items():
        stripped_def = stripped_def.replace(
    orig_name, strip_name)
    return stripped_def
```

Listing 1: S-Expression to extract the function name and change it with random values

**Dataset Creation** To prepare an instruction-based dataset for code vulnerability repair, we initially utilize the pairs of vulnerable and non-vulnerable functions extracted from MVD [66]. This resource comprises multiple C/C++ files originally sourced from NVD [67] and NIST [58]. Each file contains vulnerable code alongside its corresponding repaired version. Expanding upon this foundation, we further enrich the dataset by including *(i)* instructions, *(ii)* extracted vulnerable and repair code pair with code obfuscation, and *(iii)* vulnerability description.

**Input generation, (*I*, $C_i$).** The input consists of two components, an instruction provided in natural language and the context input, which is the code. Both elements are combined to generate the final input, which goes to the model. Based on the four tasks we created, our security experts created 20 seed instructions for each task. We created multiple duplicates of these seed instructions using GPT-4 [48] for a more robust set of instructions. The prompts are provided in Table VI in the Appendix -A.

TABLE I: CWE Counts for each vulnerable function.

| CWE Number | Count |
|---|---|
| CWE-121 | 5196 |
| CWE-134 | 3206 |
| CWE-122 | 3049 |
| CWE-124 | 1805 |
| CWE-127 | 1485 |
| CWE-78 | 1298 |
| CWE-126 | 965 |
| CWE-195 | 684 |
| CWE-194 | 24 |
| CWE-590 | 15 |
| CWE-690 | 15 |
| CWE-197 | 1 |

**Code Obfuscation.** In order to extract all function definitions and variable names from the functions, we used the S-expression of Tree-Sitter [68]. An S-expression or symbolic expression is a method to represent the Abstract Syntax Tree (AST) of source code. Each node in the tree is represented as a list, starting with the node type followed by its children, which can be terminal tokens or further nested lists. This format provides a clear and concise textual representation of the code's syntactic structure. To extract the functions and the variables, we used the following S-expression, $(function\_definition)@func-def$ in Listing 1 to replace function and variable names. We replace the user-defined function name and variable with random strings. For the same function name or the variable names, the random strings remain the same.

```
STRIP_QUERY: Final[str] = '''
(comment) @comment
'''

parser = Parser()
language = Language(str(LANGUAGES_FILE), 'cpp')
parser.set_language(language)

def remove_comments(func_def: str) -> str:
    comments = set(n.text.decode() for n,_ in
    language.query(STRIP_QUERY)
              .captures(parser.parse(func_def
    .encode()).root_node))
    new_def = func_def
    for comment in comments:
        new_def = stripped_def.replace(comment, '')
    return new_def
```

Listing 2: S-Expression to extract and remove comments

**Description Generation.** The functions extracted from MVD predominantly include comments provided by software security experts. However, these comments are often incomplete, focusing on specific statements, and some crucial vulnerable lines have multiple layers of comments. To address this, we employ tree-sitter [68] to create a method in Listing 2 for extracting comments in C/C++. Using this method, we extract the comments from within these functions. We then present the source code, stripped of comments, alongside the extracted comments to GPT-4, prompting it to generate a clear and comprehensive description of the code. Overall Table I shows the count for each of the CWEs we extracted.

> **G2 - LLM Guidelines in Selecting and Adapting Models for Security Vulnerability Detection, Description, and Repair**

*Guideline 2 (G2): Selecting and Adapting Models*
The training phase of `SecRepair` involves choosing the correct model, exposing it to data similar to what it will encounter during fine-tuning, and ensuring alignment between training and real-world application. We employ appropriate reward metrics tailored to the nature of the data, enhancing the model's learning process. This principle ensures that the trained model can effectively handle the intricacies of code repair tasks, maintaining high standards of accuracy and performance.

**Causal Decoder Model.** In order to adhere to **G2** in our proposed vulnerability analysis system, `SecRepair`, we use a unidirectional causal-decoder model called CodeLLaMa. CodeLLaMa was pre-trained to generate code based on a given instruction. Unlike regular encoder-decoder-based LLM architectures, which exhibit bidirectional properties by reading tokens forward and backward, our model focuses solely on the unidirectional aspect. It emphasizes the significance of previous tokens that are responsible for the vulnerability. By utilizing the responsible tokens for vulnerability, our causal LLM generates code repair and offers enhanced generalization capability [69]. Furthermore, causal decoder models eliminate the need for an encoder layer, resulting in faster inference times and reduced memory and energy footprint.

**Vulnerability Identification.** A prior fundamental step that needs to be completed for code vulnerability repair is vulnerability identification. Initially, we concatenate and convert the instruction I and the input function $f$ into a sequence of tokens, $t_1, t_2, ...t_p \in T$ and $y_1, y_2, ...y_q \in y$, which is the set of output tokens respectively. Then we combine both into a unique sequence $w_1, w_2, ...w_{p+q} = (t_1, ..., t_p, \$, y_1, ..., y_q)$ correspondingly. In this definition, "$" is a special token used as a separator between the input tokens into the model and the output tokens generated by the model, $p$ is the total number of input tokens, and $q$ is the total number of output tokens.

The instruction-based supervised fine-tuning method is employed to classify vulnerabilities as binary for identification purposes. The model outputs a "YES" or "NO" response indicating the presence of a vulnerability. As we utilize a generative model for vulnerability analysis, we associate each function with a specific instruction for identifying vulnerabilities. During training, the model is trained using a Cross-entropy loss function for identification purposes.

**Reinforcement Learning for Security Repair.** Given the causal decoder architecture, our model is forced to predict the next code token, and the model cannot overlook future tokens by looking at the next token during output generation. The model is provided with the input sequence $t_i \in T$ during inference, which auto-regressively generates the output, $\hat{y}_i$.

However, generating repaired code autoregressively in a supervised fine-tuning method has some challenges. In an autoregressive generative model, the model generates a token from a pool of previous tokens, which hinders the true semantic understanding of code. Furthermore, in our analysis of our proposed dataset in Table I, we see a huge disparity in the number of counts for different classes of each CWE. For instance, `CWE-121` has 5196 vulnerable and repaired code pairs, whereas `CWE-590` and `CWE-690` have only 15 examples. With such few examples for some CWE classes, it becomes extremely challenging for a supervised fine-tuning method to learn form. Therefore, we want our model to be more robust and explorative, mainly when repairing the classes of less frequent vulnerabilities in the dataset. Furthermore, regular fine-tuning only checks for cross-entropy loss, which does not enforce the model to be functionally correct. It only ensures how close the generated outcome and the ground truth are. Conclusively, a lower value of loss does not ensure the compilability nor the security correctness of code.

Therefore, to address these issues, we fine-tune our model with reinforcement learning for generative tasks where the generated code is the security-repaired version of the original input code. Furthermore, to provide comprehensive reward modeling, we use a reward function that captures the semantic and syntactic structure of the model to generate functionally correct code with proper security repairs. Using the Proximal Policy Optimization (PPO) algorithm proposed by Schulman et al., [70], which optimizes the model based on the reward it gets based on the generated repairs. We denote the input function or the vulnerable code as $f_{vul}$. The repaired output generated the model as $\hat{f_{rep}}$, and the ground truth repaired function is $f_{rep}$. We assume the repaired output sequence is $w_1^r, w_2^r, ...w_k^r$, where $k$ is the total number of output tokens of the generated repair.

We define the token-wise code generative process as a deterministic Contextual Markov Decision Process [71] with observable context only from previous tokens of vulnerable code. The repaired code sequence generated, which is the state at the $k^{th}$ token generation, is defined by our policy $\pi(.|w^r : k - 1)$, which is the probability distribution of the previous $k - 1^{th}$ input tokens from vulnerable function $f_{vul}$.

*a) Policy Optimization:* The reinforcement learning objective is to find the optimal policy by maximizing the reward metric by adding security measures while ensuring functionality is reserved. If the original vulnerable code is $f_{vul}$, the repaired code generated by the model is $\hat{f_{rep}}$, and the ground truth repaired code is $f_{rep}$. As such, we calculate the policy optimization $r_\theta$, using the following equation:

$$L(r_\theta) = log(\sigma(r_\theta(f_{vul}, f_{rep}) - r_\theta(f_{vul}, \hat{f_{rep}}))) \quad (1)$$

where $r_\theta(f_{vul}, f_{rep})$ and $r_\theta(f_{vul}, \hat{f_{rep}})$ is the scalar output of the reward model for the vulnerable code $f_{vul}$. Here $\sigma$ is an activation function, and $\theta$ is a learnable parameter.

*b) Reward:* We use BERTScore for semantic comparison using cosine similarity score to identify whether the security issues are addressed in the code. A BERT vector represents tokens that permit generating a soft similarity measure instead of exact matching since secure code can be generated. The embeddings of the generated and reference ground truth tokens are compared pairwise using cosine similarity. The embeddings contain a token's syntactic and semantic information with its neighboring tokens. As a result, during the cosine similarity comparison of the generated token with the ground truth token,

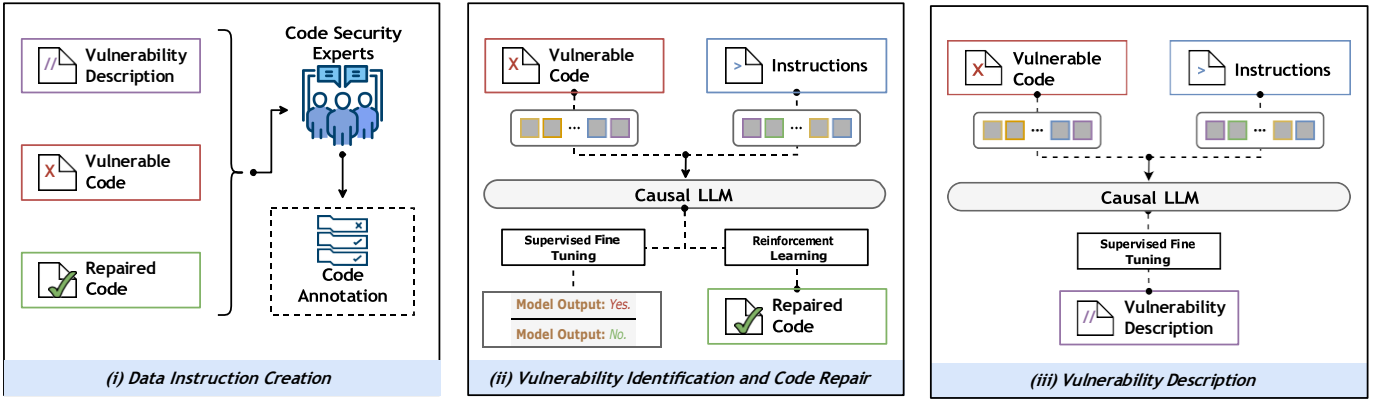

Fig. 3: `SecRepair Architecture`: The overall architecture of `SecRepair` includes *(i)* **Data Instruction Creation** involving a vulnerable code, repaired code, vulnerability description, and code security experts to annotate code; *(ii)* **Vulnerability Identification and Repair** to train a causal LLM on code coupled with instructions to identify and repair vulnerabilities using supervised fine-tuning and RL, respectively; *(iii)* **Vulnerability Description** to generate comprehensive vulnerability descriptions using supervised fine-tuning.

the synthetics, and the semantics are addressed during the training process.

The cosine similarity of a reference token from ground truth repair $t_r^i$ and a candidate token $\hat{w}_r^i$, we calculate the cosine similarity as $(t_r^i)^T \hat{w}_r^i$. Therefore, the F1 measurement of the BERTScore stands as follows;

$$R_{BERT} = \frac{1}{|t_r^i|} \sum_{t_r^i \in T} max (t_r^i)^T \hat{w}_r^i \quad (2)$$

where $R_{BERT}$ is our expected BERTScore.

**Code Vulnerability Description.** At stage 3 of Figure 3, we also fine-tune our model as a code-to-text-based format to generate a code vulnerability descriptions. We use a supervised fine-tuning method with cross-entropy loss to train the model for the description generation task.

*Guideline 3 (G3): Evaluation Procedures*

> **G3 - LLM Guidelines in Evaluation Procedures for Security Vulnerability Detection, Description, and Repair**

For `SecRepair`, the evaluation process is meticulously designed based on **G3** to ensure that the generated code is syntactically correct, with functionally security. We incorporate comprehensive evaluation metrics that assess the generated code's structural and security attributes. This holistic evaluation framework ensures that SecRepair delivers reliable and secure code repairs, addressing the limitations of existing evaluation methodologies.

We use F1, Precision, Recall, and Accuracy for vulnerability identification tasks. Since a generative model is used for vulnerability description and repair generation, metrics like BLEU, Rouge-L, Cosine Similarity, and CodeBERTScore are used. Moreover, for code vulnerability and code repair generation tasks, we use BERTScore to track how high the reward value is.

**BLEU Score.** The BLEU [64] score is a syntax-based way of evaluating machine-generated text between 0 and 1. It is a reference-based metric and captures token-based n-gram similarity.

**Rouge-L.** Similar to BLEU, Rouge-L [65] score is also a number between 0 and 1 to measure the syntax-based similarity of two generated texts. It generates a score by quantifying precision and recall by examining the longest common subsequence (LCS) between the generated and reference codes.

**Cosine Similarity.** We aim to evaluate the semantic similarity between the embeddings of LLM-generated text and the ground truth. Formally, given a set of token sequences generated by the LLM description $D = \{w_1, w_2, \ldots, w_n\}$ and the ground truth reference description $\hat{D} = \{\hat{w}_1, \hat{w}_2, \ldots, \hat{w}_n\}$, we use the sentence encoder $E$ to produce embeddings $E(D) = \{e_{w_1}, e_{w_2}, \ldots, e_{w_n}\}$ and $E(\hat{D}) = \{e_{\hat{w}_1}, e_{\hat{w}_2}, \ldots, e_{\hat{w}_n}\}$. Therefore, the full semantics of $D$ and $\hat{D}$ are represented by:

$$e_D = \frac{1}{|D|} \sum_{i=1}^{m} e_{w_i}, \quad e_{\hat{D}} = \frac{1}{|\hat{D}|} \sum_{j=1}^{n} e_{\hat{w}_j} \quad (3)$$

Therefore, we calculate the similarity score as,

$$sim(D, \hat{D}) = \frac{e_D \cdot e_{\hat{D}}^T}{\|e_D\| \cdot \|e_{\hat{D}}\|} \quad (4)$$

**CodeBERTScore.** CodeBERTScore builds on the concept of BERTScore, applying it to the domain of source code. Using CodeBERT's embeddings, it measures the semantic similarity between generated and reference codes through cosine similarity. By capturing syntactic and semantic aspects, CodeBERTScore offers an evaluation, ensuring the generated code is contextually and functionally aligned with the reference code.

TABLE II: Comparison of our Model with State-of-the-arts (SOTA) approaches in identifying vulnerabilities on our curated instruction dataset `InstructVul`.

| Training Method | Model | Acc | Pre | Rec | F1 | Acc(Vul.) | Acc(Ben) |
|---|---|---|---|---|---|---|---|
| SFT | **SecRepair** | **0.93** | **0.88** | 0.95 | 0.91 | **0.95** | **0.87** |
|  | CodeGen2 | 0.86 | 0.83 | 0.89 | 0.86 | 0.89 | 0.83 |
|  | Mistral | 0.92 | 0.87 | 0.97 | 0.92 | 0.97 | 0.86 |
|  | StarCoder | 0.88 | 1 | 0.76 | 0.86 | 0.76 | 1 |
|  | LLaMa 3 | **0.93** | **0.88** | 0.99 | **0.93** | 0.94 | 0.85 |

**Compilation Testing and Manual Evaluation.** To ensure the quality of the generated code, we use a two-step evaluation process. First, we perform compilation testing by compiling the programs using Tree-sitter, which helps us identify syntax errors and ensure that the code is syntactically correct. Next, we conduct manual evaluation by generating test cases (inputs and outputs) for ten programs and manually checking their outputs. This allows us to verify not only the functional correctness of the code but also its adherence to security standards and overall quality.

From the aforementioned evaluation metrics, we used BLEU, Rouge-L, Cosine SimilarityBy, CodeBERTScore, Compilation Testing, and Manual Evaluation to evaluate source code repair. We used BLEU, Rouge-L, Cosine Similarity, and BERTScore to describe the vulnerability. Moreover, we use Accuracy, Precision, Recall, and F1 scores for vulnerability identification tasks. Combining these evaluation methods, we ensure that the vulnerability identification, generated repaired code, and vulnerability description are reliable and high-quality.

## V. EXPERIMENTS AND DISCUSSIONS

This section uses the evaluation metrics from the previous section to answer all the research questions (RQs) we defined before.

### A. Results and Discussions

Our datasets were randomly shuffled across all of our experiments and divided into 80/10/10 for training, validation, and testing. The pretrained CodeLLaMa model with 40 layers of decoders was used to build our proposed `SercRepair`. The model was trained for three epochs with a maximum token length of 512 while the learning rate was set to $2e^{-5}$ with a batch size of 2 for our 7B parameter model. A beam size of 4 was used for the generation task, and a temperature value of 0.5 was used for optimal performance. The model was trained in a server with 8 NVIDIA A100 GPUs, each with 40 GB of memory. In the remaining part of this section, we present how the experiments were conducted to answer the three research questions mentioned in Section III.

**RQ1:** *Can we automatically identify code vulnerability of static source code?*

To ensure accurate vulnerability repair, a robust code vulnerability analysis system's primary task is to identify vulnerabilities with higher accuracy and fewer false positives and false negatives. In order to identify vulnerable code using a generative model, we compare the exact matching of the generated output. If the input code is vulnerable, then the model generates the token "YES", and otherwise, it generates the token "NO". Therefore, when the generated token "YES/NO" matches with the ground truth, it is considered a correct prediction of the vulnerability.

**Discussion.** Our instruction-based vulnerability identification system aims to determine the capability to identify vulnerable code and compare the result with the existing models. Moreover, we compare our RL technique with SFT to show the difference in the training process. We compare our proposed `SecRepair` [50] trained using SFT. We compare with 4 other state-of-the-arts LLM CodeGen2 [51], Mistral [53], StarCoder [72] and LLaMa 3 [52]. We used accuracy, precision, recall and F1 score to measure the performance of the model. Furthermore, we also tested the model on Acc (Vul) and Acc (Ben), where we only used either the vulnerable or the non-vulnerable function as the input to the model.

Table II provides a detailed comparison of our proposed model, `SecRepair`, against other state-of-the-art (SOTA) models in identifying vulnerabilities on the `InstructVul` dataset. In the SFT training method, `SecRepair` demonstrated higher performance, achieving an accuracy of 0.93, precision of 0.88, recall of 0.95, and an F1 score of 0.91. This further underscores the model's ability to learn and adapt through reinforcement learning, significantly enhancing its vulnerability detection capabilities. While other models like Mistral and LLaMa 3 also showed high performance with the RL method, `SecRepair` maintained a consistent edge, particularly in accuracy and balanced performance across vulnerable and benign classifications. The results highlight the superiority of `SecRepair` in both training paradigms, making it a reliable and efficient choice for vulnerability identification tasks.

**RQ2:** *Can we comprehensively describe the code vulnerability of the vulnerable code?*

While vulnerability identification can confirm the presence of vulnerabilities in the code, it often doesn't specify the vulnerabilities or how the vulnerability affects the code. To assist developers in properly analyzing code vulnerabilities, we aim to provide detailed descriptions of the identified vulnerabilities.

**Discussion.** Table III presents the performance of various models in generating secure code descriptions through the Supervised Fine-Tuning (SFT) training method. The models are evaluated using several metrics, including BLEU, Rouge-L, Cosine Similarity, CodeBERTScore, and BERTScore. These metrics provide a comprehensive view of each model's capability to generate accurate and semantically meaningful vulnerability descriptions.

TABLE III: Rerformance on Code Vulnerability Description by generating secure code. We use the Supervised Fine Tuning (SFT) method of training to generate vulnerability description

| Model | BLEU | Rouge-L | Cosine Similarity | BERTScore |
|---|---|---|---|---|
| **SecRepair** | 0.10 | **0.36** | **0.86** | 0.86 |
| CodeGen2 | 0.11 | 0.34 | 0.91 | **0.90** |
| Mistral | 0.12 | 0.32 | 0.90 | 0.84 |
| StarCoder | 0.08 | 0.26 | 0.86 | 0.92 |
| LLaMa 3 | 0.07 | 0.28 | 0.81 | 0.88 |

TABLE IV: Performance on code repair by generating secure code. We compare Reinforcement Learning (RL) with the Supervised Fine Tuning (SFT) method of training.

| Training | Model | BL. | Ro.-L | Cos. Sim. | CodeB.Sc. | B.Sc. |
|---|---|---|---|---|---|---|
| SFT | SecRepair | 0.76 | 0.98 | 0.85 | 0.82 | 0.91 |
| | CodeGen2 | 0.85 | 0.69 | 0.82 | 0.79 | 0.89 |
| | Mistral | 0.47 | 0.72 | 0.87 | 0.76 | 0.86 |
| | StarCoder | 0.57 | 0.90 | 0.90 | 0.75 | 0.90 |
| | LLaMa 3 | 0.68 | 0.94 | 0.91 | 0.81 | 0.92 |
| RL | SecRepair | **0.91** | **0.95** | **0.97** | 0.94 | 0.98 |
| | CodeGen2 | 0.85 | 0.69 | 0.88 | 0.90 | 0.93 |
| | Mistral | 0.97 | 0.91 | **0.97** | **0.98** | **0.99** |
| | StarCoder | 0.89 | 0.94 | 0.96 | 0.92 | 0.97 |
| | LLaMa 3 | 0.51 | 0.65 | 0.94 | 0.90 | 0.94 |

SecRepair achieves notable scores across most metrics, particularly excelling in Rouge-L and Cosine Similarity, scoring 0.36 and 0.86, respectively. This indicates that SecRepair is highly effective at generating accurate and contextually relevant descriptions. Although its BLEU score is slightly lower at 0.10, the model's strong performance in the other metrics highlights its robustness in capturing the semantic nuances necessary for secure code descriptions.

CodeGen2 demonstrates a balanced performance, leading in Cosine Similarity (0.91) and achieving the highest BERTScore (0.90). This suggests that CodeGen2 generates semantically rich and relevant descriptions, making it a strong contender for vulnerability description tasks. While slightly behind CodeGen2 in Cosine Similarity (0.90) and BERTScore (0.84), Mistral maintains competitive scores across the board, indicating consistent performance.

StarCoder and LLaMa 3, although trailing in some metrics, particularly BLEU and Rouge-L, still offer valuable insights. StarCoder's highest BERTScore of 0.92 indicates its capability to produce highly relevant and accurate descriptions despite lower scores in other metrics. Similarly, LLaMa 3, with a balanced performance and notable BERTScore of 0.88, demonstrates its potential to generate contextually appropriate descriptions.

Overall, the table highlights the strengths and weaknesses of each model in generating secure code descriptions. SecRepair and CodeGen2 emerge as the top performers, with our proposed model SecRepair excelling in most metrics and CodeGen2 showing a strong balance. The results underscore the importance of considering multiple metrics to evaluate the effectiveness of models in generating accurate and semantically meaningful vulnerability descriptions.

**RQ3:** *Can we optimize using reinforcement learning and repair the security issue in a vulnerable code?*

In order to repair the vulnerability, we trained our model with instructions to repair the vulnerability. We trained with two different processes: supervised fine-tuning and reinforcement learning.

**Discussion** The Table IV illustrates the comparative performance of various models, including our proposed model, SecRepair, evaluated on multiple metrics such as BLEU, Rouge-L, Cosine Similarity, CodeBERTScore, and BERTScore. SecRepair trained with Reinforcement Learning (RL) emerges as the best-performing model across several key metrics, achieving the highest BLEU score of 0.91, an almost perfect Rouge-L score of 0.95, and the top Cosine Similarity score of 0.97. This highlights the efficacy of the RL training method in enhancing the model's ability to generate high-quality, semantically meaningful code.

The significance of BERTScore in this context cannot be overstated. BERTScore is a semantic evaluation metric that aligns closely with human judgment by considering the contextual similarity between the generated and reference texts. The BERTScore column in the table shows a clear improvement for models trained with RL compared to those trained with Supervised Fine-Tuning (SFT). For instance, SecRepair's BERTScore increased from 0.91 with SFT to 0.98 with RL, demonstrating the model's enhanced semantic understanding and generation capabilities. This improvement underscores the importance of incorporating semantic rewards like BERTScore during the training process, as it significantly boosts the model's ability to produce code that is not only syntactically and functionally correct but also resolves security issues.

**Compilation and Manual Evaluation.** Table V provides a comprehensive evaluation of five different models: SecRepair, CodeGen2, Mistral, StarCoder, and LLaMa 3—on their ability to generate compilable code and pass security test cases under various temperature and p_value settings. The temperature parameter controls the randomness of the generated code, while the p_value influences the probability of selecting the top predictions during code generation.

Different temperature settings were used for the compilability test to generate code for ten randomly selected functions. The results were then assessed using GCC compiler to determine how many of the ten functions were compilable. SecRepair consistently showed high performance across all temperature settings, achieving 8 to 9 compilable functions out of 10. This demonstrates SecRepair's robustness in generating syntactically correct code. CodeGen2 and StarCoder also performed well, with CodeGen2 showing a slight edge at higher temperatures, reaching 9/10 compilable functions. Mistral, while showing decent performance, had a slightly lower success rate, indicating some variability in its output quality. LLaMa 3 showed consistent performance but didn't surpass SecRepair and CodeGen2.

In the manual testing column, a specific p_value of 0.5 and a temperature of 0.0 were used to generate code, followed by the generation of security test cases for the 10 functions. The outcome indicates the number of functions that passed these security tests. SecRepair once again led the pack, with 9 out of 10 functions passing the security tests, underscoring its capability to produce secure code. CodeGen2 and StarCoder followed closely, each with 8/10 functions passing the tests, demonstrating their reliability in generating secure and functional code. Mistral and LLaMa 3 had slightly lower performance, with 7/10 functions passing the tests.

```
void function_4728(){
    char * data;
    char data_buf[100] = FULL_COMMAND;
```

```
4       data = data_buf;
5       if (globalReturnsTrue()) {
6           size_t dataLen = strlen(data);
7           char * environment = GETENV(ENV_VARIABLE);
8           if (environment != NULL) {
9               strncat(data + dataLen, environment, 100
10              - dataLen - 1);
11          }
12      }
13  }
14  // Output From the Model:
15  CWE: 78 OS Command Injection Execute command in data
        using popen() Flow Variant: 11 Control flow: if
        (globalReturnsTrue()) and if(globalReturnsFalse
        ()) /This could allow attackers to execute
        unexpected, dangerous commands directly on the
        operating system. This weakness can lead to a
        vulnerability in environments in which the
        attacker does not have direct access to the
        operating system, such as in web applications.
        Alternately, if the weakness occurs in a
        privileged program, it could allow the attacker
        to specify commands that normally would not be
        accessible, or to call alternate commands with
        privileges that the attacker does not have.
```

Listing 3: **Case Study 1:** Vulnerability Description.

Overall, the results highlight our proposed `SecRepair` as the most reliable model for generating compilable and secure code across different settings. Its consistent high performance in both compilability and security testing demonstrates its effectiveness and robustness, making it a strong candidate for practical code generation tasks. CodeGen2 and StarCoder also show commendable performance, making them viable alternatives depending on specific requirements and conditions.

*B. Case Study of the Generated Outcomes*

This section provides three case studies, which will compare the outcome of our proposed `SecRepair` model.

**Case Study 1.** In this case study, we provide an analysis to demonstrate the generated vulnerability description of our generative model qualitatively. Listing 3 shows an example of the vulnerable code with a generated cybersecurity description of the code. The objective of the original code is to execute an SQL command, a standard practice in various database-based applications. However, checking the maliciousness of SQL is a critical security practice that a developer must follow to ensure the security of a program. Here, there is only a single conditional check `globalReturnsTrue()`, which is not a function of the `FULL_COMMAND` SQL code. Therefore, if malicious actors can infiltrate the `globalReturnsTrue()` function in returning true, they can essentially execute any malicious code compromising the system's security. At the bottom of Listing 3, we see the description generated by our system explaining the reasoning behind the vulnerability of the code.

**Case Study 2.** The code snippet in Listing 4 is a vulnerable function where the conditional statement `if (memcached_fatal(rc) && rc != MEMCACHED_TIMEOUT)` in line 18 is susceptible to logical errors that could lead to unexpected behavior. Specifically, the intention will reset the I/O operations `memcached_io_reset(instance)` when the return code (*rc*) indicates a fatal error that is not a timeout. If the condition is met, the following code block, intended for error handling, may not execute as expected, leaving the system in an inconsistent state.

```
1   static memcached_return_t _read_one_response(
        memcached_instance_st *instance, char *buffer,
        const size_t buffer_length, memcached_result_st
        *result) {
2     memcached_server_response_decrement(instance);
3     if (result == NULL) {
4       Memcached *root = (Memcached *) instance->root;
5       result = &root->result;
6     }
7     memcached_return_t rc;
8     if (memcached_is_binary(instance->root)) {
9       ... ...
10      rc = textual_read_one_response(instance, buffer,
         buffer_length, result);
11    }
12
13    if (memcached_fatal(rc) && rc != MEMCACHED_TIMEOUT
        ) {
14      memcached_io_reset(instance);
15    }
```

Listing 4: **Case Study 2:** I/O Operations Vulnerability Repair.

To address this vulnerability, the solution is to remove this line so that the inconsistency does not happen. However, the repair is relatively difficult to understand because the model is not given any other function for external analysis. Therefore, the SFT and RL models both removed lines 13 and 14 and repaired the vulnerability.

From the analysis of these case studies, it is seen that our model is more efficient and effective when additional lines must be added to repair the vulnerability. Our proposed RL model learns to add lines that add security measures to the code since the model was rewarded for keeping the functionality and security aspects intact. However, both models trained by both RL and SFT perform similarly when repairing a vulnerability, which includes removing some lines.

TABLE V: Code comparability testing and on ten randomly selected samples with different $p_value$, and *temperature* values. Moreover, Man. (Manual) The testing column shows how many functions pass the test cases for specific $p_value$ and *temperature* values, with manually generated test cases.

| p_value | p=0.0 | p=0.25 | p=0.50 | p=0.75 | p=1.0 | Man. Testing |
|---|---|---|---|---|---|---|
| **Model Name** | | | | | | p=0.5, t.=0.0 |
| CodeLLaMa | 8/10 | 8/10 | 9/10 | 9/10 | 7/10 | 9/10 |
| CodeGen2 | 7/10 | 8/10 | 8/10 | 9/10 | 8/10 | 8/10 |
| Mistral | 6/10 | 7/10 | 7/10 | 8/10 | 6/10 | 7/10 |
| StarCoder | 7/10 | 7/10 | 8/10 | 8/10 | 7/10 | 8/10 |
| LLaMa 3 | 7/10 | 8/10 | 8/10 | 9/10 | 7/10 | 8/10 |
| **Temperature** | t=0.0 | t=0.2 | t=0.4 | t=0.6 | t=1.0 | p=0.0, t=0.5 |
| CodeLLaMa | 8/10 | 8/10 | 9/10 | 9/10 | 7/10 | 9/10 |
| CodeGen2 | 7/10 | 8/10 | 8/10 | 9/10 | 8/10 | 8/10 |
| Mistral | 6/10 | 7/10 | 7/10 | 8/10 | 6/10 | 7/10 |
| StarCoder | 7/10 | 7/10 | 8/10 | 8/10 | 7/10 | 8/10 |
| LLaMa 3 | 7/10 | 8/10 | 8/10 | 9/10 | 7/10 | 8/10 |

**Case Study 3.** In this case study, the qualitative outcome between the SFT- and RL-based fine-tuning models is analyzed. The initial implementation of the complex algorithm exhibited a vulnerability prone to integer overflow. This flaw, residing in accumulating values within the loop, could result in unexpected behavior and potential instability when processing large datasets. The provided code snippet in Listing 5 is from the TensorFlow repository, where the code handles mutations in a TensorFlow (TF) graph during a session. This sample code is a part of our evaluation dataset. This function aims to handle mutations to a TensorFlow graph in a concurrent environment, marking sessions as un-runnable if they have already executed a specified operation. Here, the original vulnerable code excludes lines number 10 and 11. This function aims to run exclusively and exit when the loop ends. However, when the function returns, the mutex is still unlocked, blocking any execution of this function and making this function vulnerable.

A straightforward solution to this race condition problem is eliminating line number 5. However, this approach to generating the solution compromises functionality. Nevertheless, when the model undergoes fine-tuning using reinforcement learning, our proposed reward model assesses the functionality and security aspects of the input code. However, during fine-tuning, the model employs cross-entropy loss, which considers tokens at a higher level without delving deeply into the security aspects. Hence, when trained with reinforcement learning, the model preserves functionality by not eliminating line 5. Herein, the model inserts a line at line 11, introducing a statement to unlock the mutex, effectively enabling the subsequent call to execute and rectifying the vulnerability within the function.

```
void RecordMutation(TF_Graph* graph, const
    TF_Operation& op,
                    const char* mutation_type) {
  for (auto it : graph->sessions) {
    {
      mutex_lock session_lock(it.first->mu);
      if (it.first->last_num_graph_nodes > op.node.
    id()) {
        it.second = strings::StrCat(" ... ... ");
      }
    }
    // Added Repaired Line
    mutex_unlock session_unlock(it.first->mu);
  }
}
```
Listing 5: **Case Study 3:** Vulnerability Repair in Mutex Condition of a Function.

However, while adding the statement at line 11 is a probable solution, based on the training data that was available to our model, in reality, **RecordMutation** is a class that locks the mutex in instantiation and unlocks when the mutex goes out of scope. Therefore, this solution merely reflects the data it was trained on and the solution was provided solely based on the data.

## VI. Ablation Studies

In this section, we analyze the two major contributions of our work: *(i)* our proposed dataset and *(ii)* the hyperparameters of our proposed secure code generation model.

### A. Ablation Studies on Hyperparameters

This section explores the impact of two components, temperature and beam size, on generative models. Higher temperature leads to generating tokens with lower probabilistic values. Meanwhile, increasing the beam size in beam search enables tracking the top-k probable sequences and extending them with next-token probabilities.

**Temperature.** Figure 4 shows how the beam BLEU and Rouge-L score change with the change in temperature. In order to test the effect of temperature, we kept the beam search at its default value and tested the outcome with temperature values of 0.0, 0.25, 0.50, 0.75, and 1.0. We observe in Figure 4 that the scores are the lowest when the temperature is 0.0 and begin to increase when the value is at its sweet spot at 0.50. However, we again see a decrease in the score when the temperature is close to 1.

**Beam Search.** Similarly, in Figure 5, when we keep the temperature value constant and increase the beam value to 1, 2, 4, 6, and 8. We see slight improvements in the results as the beam size increases. However, this improvement comes with some drawbacks. The yellow line shows the relative time converted to a range between 0 and 1 for convenience. We see that the inference time increases almost exponentially with the increase in beam values. Here, we converted the time values within the range of 0 to 1 for the convenience of displaying them in a single graph. While the increase of beam value improves the performance, higher inference time makes this an inconvenience for the developers to use in real time.

## VII. Related Works

*a) **Advancements in Vulnerability Detection**:* Vulnerability detection methods have evolved from traditional machine learning (ML) techniques to more adaptable deep learning-based solutions with universal applicability. Lin et al. [73] emphasize the potential of ML methods for automated vulnerability discovery. However, deep learning solutions, including VulDeepecker [74], and $\mu$VulDeepecker [42] rely heavily on feature engineering for vulnerability detection. In contrast with these methods, Devign [39], VulBERTa [75] and ReVEAL [40] employ Code Property Graph-based techniques (CPG) [76] for vulnerability detection. Abstract Syntax Tree (AST)-based methods, proposed by Bilgin et al. [77] and others [78], [79], [80] retain syntactic information during detection.

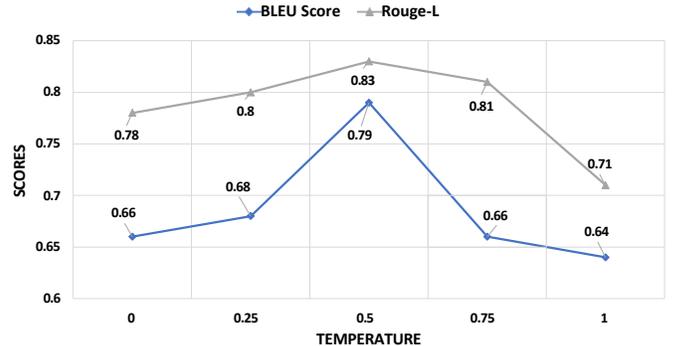

Fig. 4: **Temperature.** Ablation Study on the Performance of our Model with varying temperature.

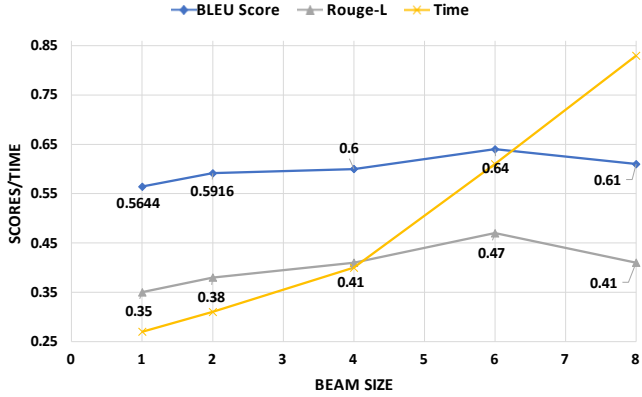

Fig. 5: **Beam Size.** Ablation Study on the Performance of our Model with Varying Beam Size.

Transformer-based models like RoBERTa [81] are also adopted by VulBERTa [75] and [82] for vulnerability detection. Islam et al. [83] propose a semantic understanding of programming languages using a combination of GCN and transformer to classify and detect vulnerabilities.

*b) Code Vulnerability with LLMs*: With the rapid progress of LLMs, their application in code development-related tasks has significantly increased. GitHub Copilot [6] has initiated a trend towards AI-assisted software development [34], [35], [36]. Notably, Codex [63] introduced a docstring-based approach where developers input instructions as docstrings, and LLMs generate corresponding code. However, most LLMs are less concerned about the security issues that come with programming languages. Pearce [23] examines LLMs' zero-shot vulnerability repairing capabilities and the bugs' challenges. Their experimental results showed that, although LLMs can generate bug fixes, they need a specifically constructed prompt to fix a particular bug. SVEN [84] proposes an adversarial technique to evaluate LLMs. They propose to generate safer code by leveraging property-specific continuous vectors to guide program generation. Moreover, [85], [86] and [87] provided an extensive study on the effectiveness of autocompletion by integrating them with different IDEs and analyzing their outcome. However, Sandoval et al. [87] suggested that LLMs help generate more functional codes with increased security issues.

*c) Vulnerability Repair*: Repairing programs poses a significant challenge as it requires identifying the vulnerable line and generating a suitable compilable line to fix the vulnerability. Previous approaches to generating vulnerability patches have involved specifying safety properties manually [88], such as preventing a program from accessing memory beyond its boundaries. Other approaches, like those proposed by Zhang et al. [89], aim to identify patch invariants or vulnerable code locations and use a set of templates to generate a repair. Recent advancements, such as Vrepair [90] and [21], employ transformer-based transfer learning methods to fix vulnerabilities in real-world programs. Similarly, VulRepair [24] utilizes pre-trained models like CodeT5 [45] and BPE tokenizer [91] for vulnerability repair. Zhang et al. [92] have further investigated the benefits and limitations of pre-trained models. With the emergence of code-based LLMs like Codex [63], recent studies by Pearce et al. [23], Jesse et al. [86], and Prenner [93] have demonstrated their ability to repair vulnerable code using zero-shot learning techniques. More recent research [94] has attempted to integrate formal verification methods with LLMs to detect security flaws in code.

*d) LLM Training and Evaluation*: Recent advancements in large language model training [52], [54] demonstrate that reinforcement learning (RL) is highly effective in guiding large language models to meet specific requirements. A recent study by Islam et al. [60] illustrates that while RL enhances code repair, traditional loss metrics perform poorly in secure code generation settings. Additionally, to evaluate generated text, metrics such as Rouge [65] and its variants like Rouge-L, METEOR [95], and chrF [96] are employed. Some scores have been adapted for the functional assessment of AI-generated code, such as CodeBLEU [97], which utilizes the graphical properties of code for comparison, and RUBY [98], which aims to consider the lexical properties of code. However, to the best of our knowledge, no method currently exists to analyze the security aspects of static code.

## VIII. CONCLUSION

In conclusion, our study offers a pioneering and in-depth exploration of the challenges in vulnerability analysis, particularly focusing on the tasks of vulnerability identification, description, and repair. We thoroughly examined and determined the security ramifications of data processing, system design, and performance analysis on generative models. Furthermore, we formulated a comprehensive threat model to find the objectives of our proposed solution. Following this, we established a set of guidelines and proposed an end-to-end solution, SecRepair, specifically designed to address the complexities associated with code vulnerabilities. This work integrates the implementation of security measures with code-generative LLMs. Central to our approach is the introduction of an instruction-based LLM, SecRepair, powered by CodeLLaMa, which is trained using reinforcement learning with a semantic reward for the security repair of source code. This innovative dataset is meticulously curated to tackle the multifaceted tasks of identifying, repairing, and describing source code vulnerabilities. Recognizing the inherent variability in generated outputs, we adopt extensive evaluation strategies, including automated techniques such as BLEU, Rouge-L, Cosine Similarity, CodeBERTScore, and test case-based evaluations, to ensure the correctness and effectiveness of the security repairs.

*A. Prompt to Generate isntructions*

Table VI illustrates the specific prompts designed to instruct a large language model (LLM) to perform various tasks related to vulnerability analysis in static source code. The table is divided into three distinct tasks: Identification, Description, and Vulnerability Repair. For each task, a detailed prompt is provided to ensure the LLM generates precise and relevant outputs without any extraneous content. For the Identification task, the LLM is guided to answer in a strict "YES/NO" format, confirming the presence of vulnerabilities. For the Description task, the LLM is prompted to describe vulnerabilities starting with their corresponding CWE numbers in the "CWE-XXX" format. For the Vulnerability Repair task, the LLM is instructed to provide only the repaired code. Each prompt set includes two example prompts to serve as templates for generating 20 different variants, ensuring comprehensive coverage and variability in instructing the LLM. These tailored prompts are crucial for optimizing the LLM's performance in generating accurate and actionable insights for software security professionals.

TABLE VI: Prompts used as an input for GPT-4 to generate 20 Instructions for each task, namely, i) Identification, ii) Description, and iii) Vulnerability Description.

| Task | Prompt |
|---|---|
| Identification | Consider yourself as a Code Security Specialist.<br>Task: You are to generate an instruction to ask a large language model to Identify Vulnerabilities in static source code.<br>Make sure that the prompt ensures the answer is in "YES/NO" format and that no extra content is generated by the LLM when using the prompt.<br>Now, create 20 different variants of these prompts.<br>Here are two sample examples:<br>1. "You are a Static Software Engineer of Code. Detect the presence of events of vulnerabilities that exist in the given code. If a vulnerability exists, answer "YES". Otherwise, answer "NO". Do not produce any extra outputs.<br>2. "As an event Vulnerability expert of static code, analyze the code to find whether any vulnerability exists in the following code.<br>If you think the code has a vulnerability, produce a single word "YES", and if vulnerability does not exist, reply "NO" only. Do not output any extra words or URLs. |
| Description | Consider yourself as a Code Security Specialist.<br>Task: You are to generate a prompt to ask a large language model to Describe Vulnerabilities in code.<br>The description is based on the CWE numbers. Make sure that the prompt ensures the answer starts with the CWE number of the vulnerability and is in "CWE-XXX" format and that no extra content is generated by the LLM when using the prompt. Here "XXX" refers to the vulnerability number.<br>Now, create 20 different variants of these prompts.<br>Here are two sample examples:<br>1. "You are a Software Engineer of Static Code. Describe the of vulnerabilities that exist in the given code.<br>Generate the vulnerability description starting with the following format only "CWE-XXX".<br>Do not produce any extra outputs.<br>2. "As an expert describing Vulnerability in static source code, analyze the code to find the vulnerability category existing in the following code. Print only the CWE number of the vulnerability in this format, "CWE-XXX" and then start the description.<br>Do not output any extra words or URLs. |
| VVulnerabilit Repair | Consider yourself as a Code Security Engineer who can repair vulnerable on static code.<br>Task: You are to generate a prompt to ask a large language model to repair a vulnerable function.<br>No extra content is generated by the LLM other than the repaired code.<br>Now, create 20 different variants of these prompts.<br>Here are two sample examples:<br>1. "You are a Software Engineer on Static Code. Repair the vulnerable function in the given static code.<br>Do not produce any extra outputs.<br>2. "As an expert repairing Vulnerability in static code, analyze the code to determine repaired code in the following code. Do not output any extra words or URLs. |